\documentclass[prl,twocolumn,showpacs,preprintnumbers,amsmath,amssymb]{revtex4}

\usepackage{mathrsfs}%hanamoji
\usepackage{graphicx}% Include figure files
\usepackage{dcolumn}% Align table columns on decimal point
\usepackage{bm}% bold math

\begin{document}

%%%%%
%\begin{minipage}{16cm}

\title{Second Law of Thermodynamics with Discrete Quantum Feedback Control}
\author{Takahiro Sagawa$^1$}
\author{Masahito Ueda$^{1,2}$}
\affiliation{$^1$Department of Physics, Tokyo Institute of Technology,
2-12-1 Ookayama, Meguro-ku, Tokyo 152-8551, Japan \\
$^2$ERATO Macroscopic Quantum Control Project, JST, 2-11-16 Yayoi, Bunkyo-ku, Tokyo 113-8656, Japan
}
\date{\today}

\begin{abstract}
A new thermodynamic inequality is derived which leads to the maximum work that can be extracted  from multi-heat baths with the assistance of discrete quantum feedback control. The maximum work is determined by the free-energy difference and a generalized  mutual information content between the thermodynamic system and the feedback controller. This maximum work can exceed that in conventional thermodynamics and, in the case of a heat cycle with two heat baths, the heat efficiency can be greater than that of the Carnot cycle.  The consistency of our results with the second law of thermodynamics is ensured by the fact that  work is needed for information processing of the feedback controller.  
\end{abstract}

\pacs{03.67.-a,05.70.Ln,05.30.-d,03.65.Ta}% PACS, the Physics and Astronomy
                             % Classification Scheme.

%\keywords{ keywords}%Use showkeys class option if keyword
                              %display desired
\maketitle
%%%%%%
%\end{minipage}\vspace{2mm}
Among a large number of studies conducted on the relationship between thermodynamics and information processing~\cite{Maxwell,Demon,Bennett,Landauer,Piechocinska,Szilard,Nielsen,Sagawa-Ueda,Maruyama,Maruyama1,Lloyd,Milburn,Scully,Kieu,Oppenheim}, particularly provoking is the work by Szilard~\cite{Szilard} who argued  that positive work $W_{\rm ext}$ can be extracted from an isothermal cycle if  Maxwell's demon plays the role of  a feedback controller~\cite{Nielsen}. It is now well understood that the role of the demon does not contradict the second law of thermodynamics, because the initialization of the demon's memory entails heat dissipation~\cite{Landauer,Bennett,Piechocinska}. We note that, in the case of an isothermal process, the second law of thermodynamics can be expressed as
\begin{equation}
W_{\rm ext} \leq - \Delta F^{\rm S},
\label{inequality0}
\end{equation}
where $\Delta F^{\rm S}$ is the difference in the Helmholtz free energy between the initial and final thermodynamic equilibrium states.

In a different context, quantum feedback control has attracted considerable attention for controlling and stabilizing a quantum system~\cite{WM1,WM2,Mabuchi1,Mabuchi2,WM3,Mabuchi3,TV}. It can be applied, for example,  to  squeezing an  electromagnetic field~\cite{WM3},  spin squeezing~\cite{Mabuchi3},  and stabilizing macroscopic coherence~\cite{TV}.  While the  theoretical framework of quantum feedback control as a stochastic dynamic system is well developed,  the possible thermodynamic gain of quantum feedback control has yet to be   fully understood.

In this Letter,  we derive a new thermodynamic inequality  which sets the  fundamental limit on the  work that can be  extracted  from  multi-heat baths with discrete quantum feedback control~\cite{Nielsen,Nielsen-Chuang},   consisting of quantum measurement~\cite{Davis-Lewis,Nielsen-Chuang} and a mechanical operation depending on the measurement outcome. The maximum work is characterized by a generalized mutual information content between the thermodynamic system and the feedback controller. We shall refer to this as the QC-mutual information content, where QC indicates that the measured system is quantal and that the measurement outcome is classical. The QC-mutual information content reduces to the classical mutual information content~\cite{Cover-Thomas} in the case of classical measurement.   In the absence of  feedback control, the new inequality~(\ref{inequality3}) reduces to the Clausius inequality.  In the case of an isothermal process, its upper bound exceeds that of  inequality~(\ref{inequality0}) by an amount proportional to the QC-mutual information content.

We consider a thermodynamic process for  system $\rm S$ which can contact  heat baths $\rm{B}_1$, ${\rm B}_2$, $\cdots$, ${\rm B}_n$ at respective temperatures $T_1$, $T_2$, $\cdots$, $T_n$.   We assume that  system $\rm S$ is in thermodynamic equilibrium in the initial and final states. For simplicity, we also assume that the initial and final temperature of $\rm S$ is given by $T \equiv (k_{\rm B} \beta)^{-1}$. This can be realized by contacting $\rm S$ with, for example,  $\rm{B}_1$ in the preparation of the initial state and during  equilibration to the final state; in this case $T=T_1$.  We do not, however, assume that the system is in  thermodynamic equilibrium between the initial and final states.  

We assume that  system $\rm S$ and  heat baths ${\rm B}_m$ are as a whole isolated and that they only come into contact with  some external mechanical systems and the feedback controller. Apart from the feedback controller, the total Hamiltonian can be written as
\begin{eqnarray}
\hat H(t) = \hat H^{\rm S} (t) + \sum_{m=1}^n ( \hat H^{{\rm SB}_m} (t) + \hat H^{{\rm B}_m}),
\label{2}
\end{eqnarray}
where $\hat H^{{\rm SB}_m} (t)$ is the interaction Hamiltonian between system $\rm S$ and heat bath ${\rm B}_m$.   The Hamiltonian $\hat H^{\rm S}(t)$ describes a mechanical operation on $\rm S$ through  such external parameters   as an applied magnetic field or  volume of the gas, and the Hamiltonian $\hat H^{{\rm SB}_m} (t)$  describes, for example,  the attachment (detachment) of an adiabatic wall or $\rm{B}_m$ to (from) $\rm S$. We consider a time evolution from  $t_{\rm i}$ to $t_{\rm f}$, assume $\hat H^{{\rm SB}_m} (t_{\rm i}) = \hat H^{{\rm SB}_m} (t_{\rm f}) = 0$ for all $m$, and write $\hat H^{\rm S}(t_{\rm i} ) = \hat H^{\rm S}_{\rm i}$ and $\hat H^{\rm S} (t_{\rm f} ) = \hat H^{\rm S}_{\rm f}$.   The time evolution of the total system with discrete quantum feedback control  can be divided into the following five stages:

\textbf{Stage 1} (\textit{Initial state}) At time $t_{\rm i}$, the initial state of $\rm S$ and that of ${\rm B}_m$ are in thermodynamic equilibrium at temperatures $T$ and  $T_m$, respectively. We assume that the density operator of the entire state is given by the canonical distribution
\begin{equation}
\hat \rho_{\rm i} = \frac{\exp (- \beta \hat H_{\rm i}^{\rm S})}{Z_{\rm i}^{\rm S}} \otimes \frac{\exp (- \beta_1 \hat H^{{\rm B}_1})}{Z^{{\rm B}_1}} \otimes \cdots \otimes \frac{\exp (- \beta_n \hat H^{{\rm B}_n})}{Z^{{\rm B}_n}}, 
\label{cani}
\end{equation}
where  $\beta_m \equiv (k_{\rm B} T_m)^{-1}$ ($m=1,2,\cdots, n$), $\ Z_{\rm i}^{\rm S} \equiv {\rm tr} \{ \exp (- \beta \hat H_{\rm i}^{\rm S}) \}$, and  $Z^{{\rm B}_m} \equiv {\rm tr} \{ \exp (- \beta_m \hat H^{{\rm B}_m}) \}$.
We denote the Helmholtz free energy of system $\rm S$ as  $F_{\rm i}^{\rm S} \equiv - k_{\rm B} T \ln Z_{\rm i}^{\rm S}$.

\textbf{Stage 2} (\textit{Unitary evolution}) From $t_{\rm i}$ to $t_1$, the entire system undergoes unitary evolution  $\hat U_{\rm i} =  {\rm T} \exp \left(  \int_{t_{\rm i}}^{t_1} \hat H (t) dt / i\hbar \right)$.

\textbf{Stage 3} (\textit{Measurement}) From $t_1$ to $t_2$, the feedback controller performs   quantum measurement on $\rm S$ described by measurement operators $\{ \hat M_k \}$ and obtains each outcome $k$ with probability $p_k$. Let $X$ be the set of outcomes $k$'s, and $\{\hat D_k \}$ be   POVM as defined by $\hat D_k \equiv \hat  M_k^\dagger \hat M_k$; we then have $p_k = {\rm tr}(\hat D_k \hat \rho)$.
We denote the pre-measurement density operator of the entire system as $\hat \rho_1$, the post-measurement density operator with outcome $k$ as $\hat \rho_2^{(k)} \equiv \hat M_k \hat \rho \hat M_k^\dagger / p_k$, and define $\hat \rho_2 \equiv \sum_k p_k \hat \rho_2^{(k)}$.  Note that our scheme can be applied not only to a quantum measurement,  but also to a classical measurement which can be described by setting $[ \hat  \rho_1 , \hat D_k ] = 0$ for all $k$.

\textbf{Stage 4} (\textit{Feedback control}) From $t_2$ to $t_3$, the feedback controller performs a mechanical operation on $\rm S$  depending on outcome $k$.  Let  $\hat U_k$ be  the corresponding unitary operator on the entire system, and $\hat \rho_3^{(k)} \equiv \hat U_k \hat \rho_2^{(k)} \hat U_k^\dagger$ be the density operator of the entire system at $t_3$ corresponding to outcome $k$. We define  $\hat \rho_3 \equiv \sum_k p_k \hat \rho_3^{(k)}$. Note that the feedback control  is characterized by  $\{\hat M_k\}$ and $\{\hat U_k \}$. 

\textbf{Stage 5} (\textit{Equilibration and final state}) From $t_3$ to $t_{\rm f}$, the entire system evolves according to unitary operator $\hat U_{\rm f}$ which is independent of outcome $k$. We assume that by $t_{\rm f}$ system $\rm S$ and heat bath $\rm {B}_m$ will have reached thermodynamic equilibrium at temperatures $T$  and $T_m$, respectively. 
We denote as $\hat \rho_{\rm f}$ the density operator of the final state of the entire system, which is related to the initial state as
\begin{equation}
\hat \rho_{\rm f} =\mathcal E(\hat \rho_{\rm i}) \equiv \sum_k \hat U_{\rm f}\hat U_k \hat M_k \hat U_{\rm i} \hat \rho_{\rm i} \hat U_{\rm i}^\dagger \hat M_k^\dagger \hat U_k^\dagger \hat U_{\rm f}^\dagger.
\end{equation}
We emphasize that $\hat \rho_{\rm f}$ need not equal the rigorous canonical distribution $\hat \rho^{\rm can}_{\rm f}$, as given by
\begin{equation}
\hat \rho_{\rm f}^{\rm can} = \frac{\exp (- \beta \hat H_{\rm f}^{\rm S})}{Z_{\rm f}^{\rm S}} \otimes \frac{\exp (- \beta_1 \hat H^{{\rm B}_1})}{Z^{{\rm B}_1}} \otimes \cdots \otimes \frac{\exp (- \beta_n \hat H^{{\rm B}_n})}{Z^{{\rm B}_n}},
\label{canf}
\end{equation}
where  $Z_{\rm f}^{\rm S} \equiv  {\rm tr} \{ \exp (- \beta \hat H_{\rm f}^{\rm S}) \}$.  We only assume that the final state is in thermodynamic equilibrium from a macroscopic point of view~\cite{Sagawa-Ueda}.

We will proceed to our  main analysis.  The difference in the von Neumann entropy between the initial and final states can be bounded from the foregoing analysis as follows:
\begin{equation}
\begin{split}
&S(\hat \rho_{\rm i})-S(\hat \rho_{\rm f}) \\
=& S(\hat \rho_1) - S(\hat \rho_3) \\
\leq& S(\hat \rho_1) - \sum_k p_k S(\hat \rho_3^{(k)}) \\
=& S(\hat \rho_1) - \sum_k p_k S(\hat \rho_2^{(k)}) \\
=& S(\hat \rho_1) + \sum_k {\rm tr} \left( \sqrt{\hat D_k}\rho_1\sqrt{\hat D_k} \ln \frac{\sqrt{\hat D_k}\hat \rho_1\sqrt{\hat D_k}}{p_k} \right) \\
=& S(\hat \rho_1) + H(\{ p_k \}) + \sum_k {\rm tr} ( \sqrt{\hat D_k}\rho_1\sqrt{\hat D_k} \ln \sqrt{\hat D_k}\rho_1\sqrt{\hat D_k}),
\end{split}
\label{main-inequality}
\end{equation}
where $S(\hat \rho) \equiv -{\rm tr} (\hat \rho \ln \hat \rho)$ is the von Neumann entropy and $H(\{ p_k \}) \equiv -\sum_{k \in X} p_k \ln p_k$ is the Shannon information content.  Note that in deriving the inequality~(\ref{main-inequality}), we used the convexity of the von Neumann entropy, i.e. $S(\sum_k p_k \hat \rho_3^{(k)}) \geq \sum_k p_k S(\hat \rho_3^{(k)})$.
Defining notations $\tilde{H} (\hat \rho_1,X) \equiv - \sum_k {\rm tr}(\sqrt{\hat D_k}\hat \rho_1\sqrt{\hat D_k} \ln \sqrt{\hat D_k}\hat \rho_1\sqrt{\hat D_k})$ and
\begin{equation}
I(\hat \rho_1\! : \! X) \equiv S(\hat \rho_1) + H(\{ p_k \}) - \tilde{H} (\hat \rho_1,X),
\end{equation}
we obtain
\begin{equation}
S(\hat \rho_{\rm i}) - S(\hat \rho_{\rm f}) \leq I(\hat \rho_1\! : \! X).
\label{inequality1}
\end{equation}

We refer to $I(\hat \rho_1\! : \! X)$ as the QC-mutual information content  which describes the information about the measured system that has been obtained by measurement.    As shown later, $I(\hat \rho_1\! : \! X)$ satisfies 
\begin{equation}
0 \leq I(\hat \rho_1\! : \! X) \leq H(\{ p_k \}).
\label{QC}
\end{equation}
We note that $I(\hat \rho_1\! : \! X) = 0$ holds for all state $\hat \rho_1$ if and only if $\hat D_k$ is proportional to the identity operator for all $k$, which means that we cannot obtain any information about the system by this measurement.  On the other hand, $I(\hat \rho_1\! : \! X) = H(\{ p_k \})$  holds  if and only if $\hat D_k$  is the projection operator satisfying $[\hat \rho_1, \hat D_k] = 0$ for all $k$, which means that the measurement on state $\hat \rho_1$ is classical and error-free.  
In the case of classical measurement  (i.e.  $[\hat \rho_1, \hat D_k ] = 0$ for all $k$),  $I(\hat \rho_1\! :\! X)$ reduces to the classical mutual information content.  In fact, we can write $I(\hat \rho_1\! :\! X)$ in this case as 
$I(\hat \rho_1\! :\! X) = -\sum_i q_i \ln q_i -  \sum_{k,i} q_i p(k|i) \ln p(k|i)$,
where $\hat \rho_1 \equiv \sum_i q_i | \psi_i \rangle \langle \psi_i |$ is the spectrum decomposition of the measured state, and $p(k|i) \equiv \langle \psi_i | \hat D_k | \psi_i \rangle$ can be interpreted as the conditional probability of obtaining outcome  $k$ under the condition that the measured state is $| \psi_i \rangle$.

$I(\hat \rho_1 \! : \! X)$ can be written as $I(\hat \rho_1 \! : \! X) = \chi (\{ \hat \rho_2^{(k)} \} ) - \Delta S_{\rm meas}$, where $\chi (\{\hat \rho_2^{(k)} \} ) \equiv S(\hat \rho_2) - \sum_{k \in X} p_k S(\hat \rho_2^{(k)})$ is the Holevo $\chi$ quantity  which sets the Holevo bound~\cite{Nielsen-Chuang,Holevo}, and  $\Delta S_{\rm meas} \equiv S(\hat \rho_2) - S(\hat \rho_1)$ is the difference in the von Neumann entropy between the pre-measurement and post-measurement states.    If  $\Delta S_{\rm meas}=0$ holds, that is, if the measurement process does not disturb the measured system, then $I(\hat \rho_1 \! : \! X)$ reduces to the Holevo $\chi$ quantity;  in this case, the upper bound of the entropy reduction with   discrete quantum feedback control  is given by the distinguishability of post-measurement states $\{ \hat \rho_2^{(k)} \}$.

Nielsen \textit{et al}. have derived inequality $S(\hat \rho_{\rm i})- S(\hat \rho_{\rm f}) \leq S(\hat \rho_{\rm i},\mathcal E)$ \cite{Nielsen,Nielsen-Chuang}, where $S( \hat \rho_{\rm i},\mathcal E)$ is the entropy exchange which depends on  entire process $\mathcal E$, including the feedback process.  In contrast,  our inequality~(\ref{inequality1}) is bounded by $I(\hat \rho_1\! :\! X)$ which  does not depend on the feedback process, but only  depends on pre-measurement state $\hat \rho_1$ and  POVM $\{ \hat D_k \}$, namely, on the information gain by the measurement alone.

It follows from inequality (\ref{inequality1}) and Klein's inequality~\cite{Tasaki} that
\begin{equation}
S(\hat \rho_{\rm i} ) \leq -{\rm tr} (\hat \rho_{\rm f} \ln \hat \rho_{\rm f}^{\rm can}) + I(\hat \rho_1\! : \! X).
\label{inequality2}
\end{equation}
Substituting Eqs.~(\ref{cani}) and (\ref{canf}) into inequality (\ref{inequality2}), we have
\begin{equation}
(\! E_{\rm i}^{\rm S} - E_{\rm f}^{\rm S} \! )  + \! \sum_{m=1}^n \! \frac{T}{T_m}  (\! E_{\rm i}^{{\rm B}_m} - E_{\rm f}^{{\rm B}_m}\!) \! \leq \!  F_{\rm i}^{\rm S} - F_{\rm f}^{\rm S}\! + k_{\rm B}TI(\! \hat \rho_1\! :\! X \!),
\end{equation}
where  $E_{\rm i}^{\rm S} \equiv {\rm tr}(\hat H_{\rm i}^{\rm S} \hat \rho_{\rm i})$, $E_{\rm f}^{\rm S} \equiv {\rm tr}(\hat H_{\rm f}^{\rm S} \rho_{\rm f})$, $E_{\rm i}^{{\rm B}_m} \equiv {\rm tr}(\hat H^{{\rm B}_m} \hat \rho_{\rm i})$, and  $E_{\rm f}^{{\rm B}_m} \equiv {\rm tr}(\hat H^{{\rm B}_m} \hat \rho_{\rm f})$. Defining the difference in the internal energy between the initial and final states of system $S$ as $\Delta U^{\rm S} \equiv E_{\rm f}^{\rm S} - E_{\rm i}^{\rm S}$,   the heat exchange between system $\rm S$ and heat bath ${\rm B}_m$ as $Q_m \equiv E_{\rm i}^{{\rm B}_m} - E_{\rm f}^{{\rm B}_m}$, and  the difference in the Helmholtz free energy of system $\rm S$  as $\Delta F^{\rm S} \equiv F_{\rm f}^{\rm S} - F_{\rm i}^{\rm S}$,  we obtain
\begin{equation}
- \Delta U^{\rm S} + \sum_{m=1}^n \frac{T}{T_m} Q_m \leq - \Delta F^{\rm S} + k_{\rm B}TI(\hat \rho_1\! :\! X).
\label{inequality3}
\end{equation}
This is the main result of this Letter. Inequality (\ref{inequality3}) represents the second law of thermodynamics in the presence of a discrete quantum feedback control, where the effect of the feedback control is described by the last term.  For a thermodynamic heat cycle in which  $I(\hat \rho_1\! :\! X)=0$, $\Delta U^{\rm S}=0$, and $\Delta F^{\rm S} = 0$ hold, inequality (\ref{inequality3})  reduces to the Clausius inequality
\begin{equation}
\sum_{m=1}^n \frac{Q_m}{T_m}  \leq 0.
\end{equation}
The equality in (\ref{inequality3}) holds if and only if $\hat \rho_3^{(k)}$ is independent of  measurement outcome $k$ (i.e.  the feedback control is perfect), and  $\hat \rho_{\rm f}$ coincides with $\hat \rho_{\rm f}^{\rm can}$.

We will discuss two important cases for inequality.  Let us first consider a situation in which the system undergoes an isothermal process in contact with  single heat bath $\rm B$ at temperature $T$.  In this case,  (\ref{inequality3}) reduces to
\begin{equation}
W_{\rm ext} \leq - \Delta F^{\rm S} + k_{\rm B}TI(\hat \rho_1\! :\! X),
\label{inequality4}
\end{equation}
where the first law of thermodynamics, $W_{\rm ext} = \sum_{m=1}^n Q_m - \Delta U^{\rm S}$, is used.
Inequality (\ref{inequality4}) implies  that we can extract work greater than $- \Delta F^{\rm S}$ from a single heat bath with  feedback control, but that we cannot extract work larger than $- \Delta F^{\rm S} + k_{\rm B}TI(\hat \rho_1\! :\! X)$.
If we do not get any information, (\ref{inequality4}) reduces to (\ref{inequality0}). On the other hand, in the case of  classical and error-free measurement,  (\ref{inequality4})  becomes $W_{\rm ext} \leq - \Delta F^{\rm S} + k_{\rm B} T H(\{ p_k \})$.

The upper bound of inequality~(\ref{inequality4}) can be achieved with the Szilard engine~\cite{Szilard} which is described as follows.  A molecule is initially in thermal equilibrium in a box in contact with a heat bath at temperature $T$. We quasi-statically partition the  box into two smaller boxes of equal volume, and perform a measurement on the system to find out in which box the molecule is.  When the molecule is found in the right one, we remove the left one and move the right one to the left position, which is the feedback control.  We then expand the box quasi-statically and isothermally so that the final state of the entire system returns to the initial state from a macroscopic point of view. During the entire process, we obtain $\ln 2$ of information and extract $k_{\rm B}T \ln 2$ of work from the system.

We next consider a heat cycle which contacts two heat baths: $\rm B_{\rm H}$ at temperature $T_{\rm H}$ and $\rm B_{\rm L}$ at  $T_{\rm L}$ with $T_{\rm H}> T_{\rm L}$.  We assume that  $\hat H_{\rm i}^{\rm S} = \hat H_{\rm f}^{\rm S}$, $\Delta U^{\rm S} = 0$, and $\Delta F^{\rm S} = 0$.  Noting that $W_{\rm ext} = Q_{\rm H} + Q_{\rm L}$, we can obtain
\begin{equation}
W_{\rm ext} \leq \left( 1- \frac{T_{\rm L}}{T_{\rm H}}\right) Q_{\rm H} +  k_{\rm B}T_{\rm L}I(\hat \rho_1\! :\! X).
\label{inequality5}
\end{equation}
Without a feedback control,  (\ref{inequality5}) shows that  the upper bound for the efficiency of heat cycles is given by that of the Carnot cycle: $W_{\rm ext}/Q_{\rm H} \leq 1-T_{\rm L}/T_{\rm H}$.  With feedback control,   (\ref{inequality5}) implies that the upper bound for the efficiency of  heat cycles becomes larger than that of the Carnot cycle.  The upper bound of~(\ref{inequality5}) can be achieved by performing a Szilard-type operation during the isothermal process of the one-molecule Carnot cycle; if we perform the measurement and feedback with $\ln 2$ of information in the same scheme as the Szilard engine during the isothermal process at temperature $T_{\rm H}$, the work that can be extracted is given by $W_{\rm ext} = (1-T_{\rm L}/T_{\rm H})(Q_{\rm H} - k_{\rm B}T_{\rm H} \ln 2) + k_{\rm B}T_{\rm H} \ln 2 = (1-T_{\rm L}/T_{\rm H})Q_{\rm H} + k_{\rm B}T_{\rm L} \ln 2$. Note that we can reach the same bound by performing the Szilard-type operation during the isothermal process at temperature $T_{\rm L}$.

We now  prove  inequality~(\ref{QC}).  For simplicity of notation, we consider a quantum system denoted as $\rm Q$ in general, instead of $\rm S$ and ${\rm B}_m$'s.  The measured state of system $\rm Q$ is written as $\hat \rho$, and POVM as $\{ \hat D_k \}_{k \in X}$.  We introduce auxiliary system $\rm R$ which is spanned by orthonormal basis $\{ | \phi_k \rangle \}_{k \in X}$, and  define two states $\hat \sigma_1$ and $\hat \sigma_2$ of $\rm{Q}+\rm{R}$ as $\hat \sigma_1 \equiv \sum_k \sqrt{\hat \rho} \hat D_k \sqrt{\hat \rho} \otimes | \phi_k \rangle \langle \phi_k|$ and $\hat \sigma_2 \equiv \sum_k \sqrt{\hat D_k} \hat \rho \sqrt{\hat D_k} \otimes | \phi_k \rangle \langle \phi_k |$.  It can be  shown that ${\rm tr}(\sqrt{\hat \rho} \hat D_k \sqrt{\hat \rho})={\rm tr}(\sqrt{\hat D_k} \hat \rho \sqrt{\hat D_k})=p_k$, ${\rm tr}_{\rm R}(\hat \sigma_1) = \hat \rho$, and ${\rm tr}_{\rm Q} (\hat \sigma_1) = \sum_k p_k | \phi_k \rangle \langle \phi_k| \equiv \hat \rho_{\rm R}$.  Defining $\hat \sigma_1^{(k)} \equiv \sqrt{\hat \rho} \hat D_k \sqrt{\hat \rho}/p_k$, $\hat \sigma_2^{(k)} \equiv \sqrt{\hat D}_k \hat \rho \sqrt{\hat D_k}/p_k$ and $\hat \rho' \equiv \sum_k p_k \hat \sigma_2^{(k)}$,  we have
\begin{equation}
\begin{split}
S(\hat \sigma_2) &= \sum_k p_k S\left(\sqrt{\hat D_k} \hat \rho \sqrt{\hat D_k}\otimes | \phi_k \rangle \langle \phi_k | /p_k \right) + H(\{ p_k \}) \\
&= \sum_k p_k S(\hat \sigma_2^{(k)}) +H(\{ p_k \}) =\tilde{H}(\hat \rho,X).
 \end{split}
 \label{QC2}
\end{equation}
Since $S(\hat L^\dagger \hat L) = S(\hat L \hat L^\dagger)$ holds for any linear operator $\hat L$, we have $S(\hat \sigma_2) =  \sum_k p_k S(\hat \sigma_2^{(k)}) +H(\{ p_k \}) =  \sum_k p_k S(\hat \sigma_1^{(k)}) +H(\{ p_k \}) = S(\hat \sigma_1)$. 
Therefore 
\begin{equation}
\tilde{H}(\hat \rho,X)=S(\hat \sigma_1) \leq S(\hat \rho)+S(\hat \rho_{\rm R}) = S(\hat \rho) +H(\{ p_k \}),
\label{QC3}
\end{equation}
which implies $I(\hat \rho \! : \! X) \geq 0$. The equality in (\ref{QC3}) holds for all $\hat \rho$ if and only if $\hat \sigma_1$ can be written as tensor product $\hat \rho \otimes \hat \rho_{\rm R}$ for all $\hat \rho$: that is, $\hat D_k$ is proportional to the identity operator for all $k$.    
We will next show that $I(\hat \rho \! : \! X) \leq H(\{ p_k \})$.  We make spectral decompositions as $\hat \rho = \sum_i q_i | \psi_i \rangle \langle \psi_i |$ and  $\hat \rho' = \sum_j r_j | \psi_j' \rangle \langle \psi_j' |$, where  $r_j = \sum_i q_i d_{ij} $, and define $d_{ij} \equiv \sum_k | \langle \psi_i | \sqrt{\hat D_k} | \psi_j' \rangle |^2$, where $\sum_i d_{ij}=1$ for all $j$ and $\sum_j d_{ij}=1$ for all $i$.  It follows from the convexity of  $-x\ln x$ that
$S(\hat \rho) = - \sum_i q_i \ln q_i \leq -\sum_j r_j \ln r_j = S(\hat \rho')$.  Therefore,
\begin{equation}
\begin{split}
H(\{ p_k \})-I(\hat \rho \! : \! X) &=  \tilde{H}(\hat \rho,X) - S(\hat \rho) \\
&= H(\{ p_k \}) + \sum_k p_k S(\hat \sigma_2^{(k)}) - S(\hat \rho) \\
&\geq H(\{ p_k \}) + \sum_k p_k S(\hat \sigma_2^{(k)}) - S(\hat \rho') \\
&\geq 0.
\label{QC4}
\end{split}
\end{equation}
It can be shown that the left-hand side is equal to zero  for all $\hat \rho$ if and only if $\hat D_k$ is a projection operator satisfying $[\hat \rho, \hat D_k] = 0$ for all $k$.

Our results do not contradict the  second law of thermodynamics, because there exists an energy cost for information processing of the feedback controller~\cite{Landauer,Bennett,Piechocinska}.  Our results are independent of the state  of the feedback controller, be it in thermodynamic equilibrium or not, because the feedback control is  solely characterized by  $\{\hat M_k \}$ and $\{\hat U_k\}$.

In conclusion, we have extended the  second law of thermodynamics to a situation in which a general thermodynamic process is accompanied by discrete quantum feedback control.  We have applied  our main result (\ref{inequality3}) to an isothermal process and a heat cycle with two heat baths, and respectively obtained inequalities~(\ref{inequality4}) and~(\ref{inequality5}).  We have identified the maximum work that can be extracted from a heat bath(s) with  feedback control; the maximum work is characterized by the generalized mutual information content between the measured system and the feedback controller.   

%%%%%%%%%%%%%%%%%%%%%%%%%%%%%%Macro 1%%%%%%%%%%%%%%%%%%%%%%%%%%%%%%%%%%%%%%%%%%%%%%%%%%%%%%%%%%%%%%%%%%%%%%%%%%%%%%%%%%%

\begin{acknowledgments}
This work was supported by a Grant-in-Aid for Scientific Research (Grant No.\ 17071005) and by a 21st Century COE program at Tokyo Tech, ``Nanometer-Scale Quantum Physics'', from the Ministry of Education, Culture, Sports, Science and Technology of Japan.
\end{acknowledgments}

%\newpage %Just because of unusual number of tables stacked at end
%\bibliography{References}% Produces the bibliography via BibTeX.

\end{document}